\documentclass[a4paper]{article}

\usepackage{INTERSPEECH2021}
\usepackage{subcaption}
\usepackage{amsmath}
\usepackage{mathtools}
\usepackage{cite}

\title{Diff-TTS: A Denoising Diffusion Model for Text-to-Speech}
\name{Myeonghun Jeong$^1$, Hyeongju Kim$^2$, Sung Jun Cheon$^1$, Byoung Jin Choi$^1$, and Nam Soo Kim$^1$}
\address{
  $^1$Department of Electrical and Computer Engineering and INMC,
Seoul National University, Seoul, South Korea\\
  $^2$Neosapience, Inc., Seoul, South Korea}
\email{mhjeong@hi.snu.ac.kr, hyeongju@neosapience.com, \{sjcheon, bjchoi\}@hi.snu.ac.kr nkim@snu.ac.kr}

\begin{document}

\maketitle
\begin{abstract}
Although neural text-to-speech~(TTS) models have attracted a lot of attention and succeeded in generating human-like speech, there is still room for improvements to its naturalness and architectural efficiency. In this work, we propose a novel non-autoregressive TTS model, namely Diff-TTS, which achieves highly natural and efficient speech synthesis. Given the text, Diff-TTS exploits a denoising diffusion framework to transform the noise signal into a mel-spectrogram via diffusion time steps. In order to learn the mel-spectrogram distribution conditioned on the text, we present a likelihood-based optimization method for TTS. Furthermore, to boost up the inference speed, we leverage the accelerated sampling method that allows Diff-TTS to generate raw waveforms much faster without significantly degrading perceptual quality. Through experiments, we verified that Diff-TTS generates 28 times faster than the real-time with a single NVIDIA 2080Ti GPU.

\end{abstract}

\section{Introduction}

Most of deep learning-based speech synthesis systems consist of two parts: (1) a text-to-speech~(TTS) model that converts text into an acoustic feature such as mel-spectrogram~\cite{wang2017tacotron,shen2018natural,ping2017deep,lee2020bidirectional} and (2) a vocoder that generates time-domain speech waveform using this acoustic feature~\cite{prenger2019waveglow,kong2020hifi,oord2016wavenet,ping2018clarinet}. In this paper, we address the acoustic modeling of neural TTS.

TTS models have attracted a lot of attention in recent years due to the advance in deep learning.
Modern TTS models can be categorized into two groups according to their formulation: (i) autoregressive (AR) models and (ii) non-AR models. The AR models can generate high-quality samples by factorizing the output distribution into a product of conditional distribution in sequential order.
The AR models such as Tacotron2 and Transformer-TTS produce natural speech, but one of their critical limitations is that inference time increases linearly with the length of mel-spectrograms~\cite{shen2018natural,li2019neural}. Furthermore, the AR model lacks robustness in some cases, e.g., word skipping and repeating, due to accumulated prediction error.
Recently, various non-AR TTS models have been proposed to overcome the shortcomings of the AR models~\cite{valle2020flowtron, peng2020non,lancucki2020fastpitch, miao2020efficienttts, donahue2020end}. Although non-AR TTS models are capable of synthesizing speech in a stable way and their inference procedure is much faster than AR models, the non-AR TTS models still have some limitations. For instance, since feed-forward models such as FastSpeech1,2~\cite{ren2019fastspeech,ren2020fastspeech} and Speedyspeech~\cite{vainer2020speedyspeech} cannot produce diverse synthetic speech since they are optimized by a simple regression objective function without any probabilistic modeling. Additionally, the flow-based generative models such as Flow-TTS~\cite{miao2020flow} and Glow-TTS~\cite{kim2020glow} are parameter-inefficient due to architectural constraints imposed on normalizing flow-based models.

Meanwhile, another generative framework called denoising diffusion has shown state-of-the-art performance on image generation, and raw audio synthesis~\cite{ho2020denoising,xiao2020vaebm,nichol2021improved,chen2020wavegrad,kong2020diffwave}. The denoising diffusion models can be stably optimized according to maximum likelihood and enjoy the freedom of architecture choices. 

In light of the advantages of denoising diffusion, we propose Diff-TTS, a novel non-AR TTS
that achieves robust, controllable, and high-quality speech synthesis. To train Diff-TTS without any auxiliary loss function, we present a log-likelihood-based optimization method based on denoising diffusion framework for TTS. In addition, since Diff-TTS has Markov-chain constraint, which accompanies slow inference speed, we also introduce the accelerated sampling method demonstrated in~\cite{song2020denoising}. 
Through the experiments, we show that Diff-TTS provides an efficient and powerful synthesis system compared to other best-performing models.
The contributions of our work are as follows:
\begin{itemize}
\item To the best of our knowledge, it is the first time that a denoising diffusion probabilistic model~(DDPM) was applied to non-AR TTS. The Diff-TTS can be stably trained without any constraint on model architecture.
\item We show that Diff-TTS generates high fidelity audios in terms of the Mean Opinion Score~(MOS) with half parameters of the Tacotron2 and Glow-TTS.
\item By applying the accelerated sampling, Diff-TTS allows users to control the trade-off between sample quality and inference speed based on a computational budget available on the device. Furthermore, Diff-TTS achieves a significantly faster inference speed than real-time.
\item We analyze how Diff-TTS controls speech prosody. The pitch variability of Diff-TTS depends on variances of latent space and additive noise. Diff-TTS can effectively control the pitch variability by multiplying a temperature term to variances.
\end{itemize}

This paper is organized as follows: we first introduce the denoising diffusion framework for TTS. Then, we present the simple architecture of  Diff-TTS. In the experiments, we analyze a number of advantages of Diff-TTS compared to other state-of-the-art TTS models.

\begin{figure}[ht]
  \centering
  \includegraphics[width=\linewidth]{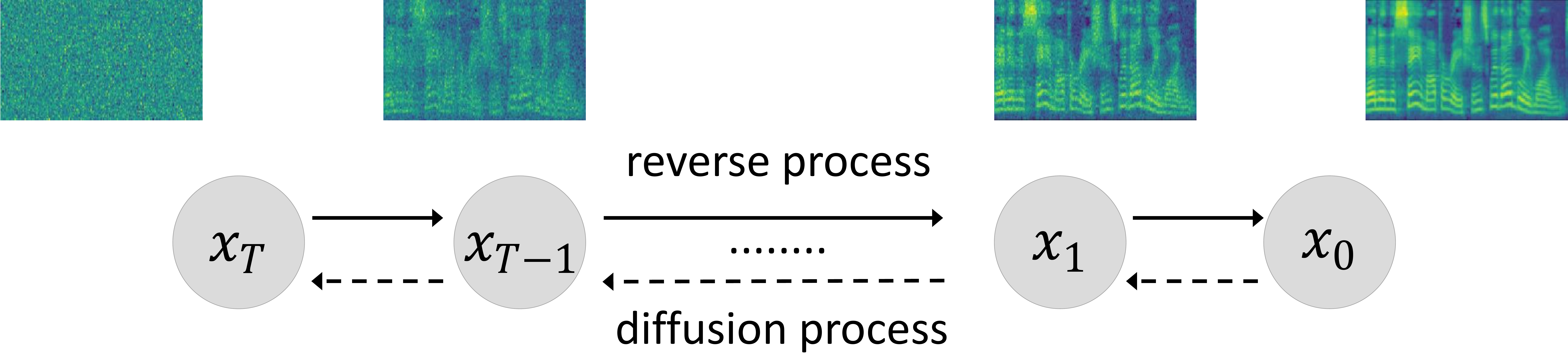}
  \caption{Graphical model for the reverse process and the diffusion process.}
  \label{fig:ddpm_graphical}
\end{figure}
\section{Diff-TTS}
In this section, we first present the probabilistic model of Diff-TTS for a mel-spectrogram generation. Then we formulate a likelihood-based objective function for training Diff-TTS. Next, we introduce two kinds of sampling methods for synthesis: (1) normal diffusion sampling based on the Markovian diffusion process and (2) accelerated version of diffusion sampling. Finally, we present the model architecture of Diff-TTS for high fidelity synthesis.

\subsection{Denoising diffusion model for TTS}
Diff-TTS converts the noise distribution to a mel-spectrogram distribution corresponding to the given text. As shown in Fig.~\ref{fig:ddpm_graphical}, the mel-spectrogram is gradually corrupted with Gaussian noise and transformed into latent variables. This process is termed the \textit{diffusion process}. Let $x_1,\dots,x_T$ be a sequence of variables with the same dimension where $t=0,1,\dots,T$ is the index for diffusion time steps.
Then, the diffusion process transforms mel-spectrogram $x_0$ into a Gaussian noise $x_T$ through a chain of Markov transitions. Each transition step is predefined with a variance schedule $\beta_1, \beta_2, ..., \beta_T$. More specifically, each transformation is performed according to the Markov transition probability $q(x_t|x_{t-1},c)$ assumed to be independent of the text $c$ and it is defined as follows:
\begin{equation}
  q(x_t|x_{t-1},c) = \mathcal{N}(x_t;\sqrt{1-\beta_t}x_{t-1},\beta_tI).
  \label{eq: diffusion transition kernel}
\end{equation}
The whole diffusion process $q(x_{1:T}|x_0,c)$ is the Markov process and can be factorized as follows:
\begin{equation}
  q(x_1\dots ,x_T|x_0,c) = \prod_{t=1}^{T} q(x_t|x_{t-1}).
  \label{eq2: DDPM markov chain assumption}
\end{equation}
The \textit{reverse process} is a mel-spectrogram generation procedure that is exactly backward of the diffusion process. Unlike the diffusion process, the goal of the reverse process is to recover a mel-spectrogram from Gaussian noise. The reverse process is defined as the conditional distribution $p_{\theta}(x_{0:T-1}|x_T,c)$, and it can be factorized into multiple transitions based on Markov chain property:
\begin{equation}
  p_\theta(x_0\dots ,x_{T-1}|x_T,c) = \prod_{t=1}^{T} p_\theta(x_{t-1}|x_{t},c).
  \label{eq1: DDPM markov chain assumption}
\end{equation}
Through the reverse transitions $p_\theta(x_{t-1}|x_t,c)$, the latent variables are gradually restored to a mel-spectrogram corresponding to the diffusion time-step with text condition. In other words, Diff-TTS learns a model distribution $p_{\theta}(x_0|c)$ obtained from the reverse process. Let $q(x_0|c)$ be the mel-spectrogram distribution. 
In order for the model to approximate $q(x_0|c)$ well, the reverse process aims to maximize the log-likelihood of the mel-spectrogram: $\mathbb{E}_{logq(x_0|c)}[logp_{\theta}(x_0|c)]$. Since $p_{\theta}(x_0|c)$ is intractable, we utilize the parameterization trick which is demonstrated in~\cite{ho2020denoising} to calculate the variational lower bound of the log-likelihood in a closed form.
Let $\alpha_t = 1-\beta_t$ and $\bar{\alpha}_{t} = \Pi^{t}_{t'=1}\alpha_{t'}$. The training objective of Diff-TTS is as follows:
\begin{equation}
  \underset{\theta}{min}L(\theta) = \mathbb{E}_{x_0,\epsilon,t} \\
  \Arrowvert \epsilon - \epsilon_{\theta}(\sqrt{\bar{\alpha_t}}x_0 + \sqrt{1-\bar{\alpha}_t}\epsilon,t,c) \Arrowvert_1,
  \label{eq4: DDPM Training objective}
\end{equation}
where $t$ is uniformly taken from the entire diffusion time-step.
Diff-TTS does not need any auxiliary losses except L1 loss function between the output of model $\epsilon_{\theta}(\cdot)$ and Gaussian noise $\epsilon \sim \mathcal{N}(0,I)$.
During inference, Diff-TTS retrieves a mel-spectrogram from a latent variable by iteratively predicting the diffusing noise added at each forward transition with $\epsilon_{\theta}(x_t, t, c)$ and removing the corrupted part as follows:
\begin{equation}
  x_{t-1} = \frac{1}{\sqrt{\alpha_t}}(x_t-\frac{1-\alpha_t}{\sqrt{1-\bar{\alpha}_t}}\epsilon_\theta(x_t,t,c)) + \sigma_tz_t,
  \label{eq5: DDPM Sampling objective}
\end{equation}
where $z_t\sim\mathcal{N}~(0,I)$ and  $\sigma_t=\eta\sqrt{\frac{1-\bar{\alpha}_{t-1}}{1-\bar{\alpha}_t}\beta_t}$. The temperature term $\eta$ is a scaling factor of the variance. Note that diffusion time-step $t$ is also used as an input to Diff-TTS, which allows shared parameters for all diffusion time-steps. As a result, the final mel-spectrogram distribution $p(x_0|c)$ is obtained through iterative sampling over all of the preset time steps.

\begin{figure}[t]
  \centering
  \includegraphics[width=\linewidth]{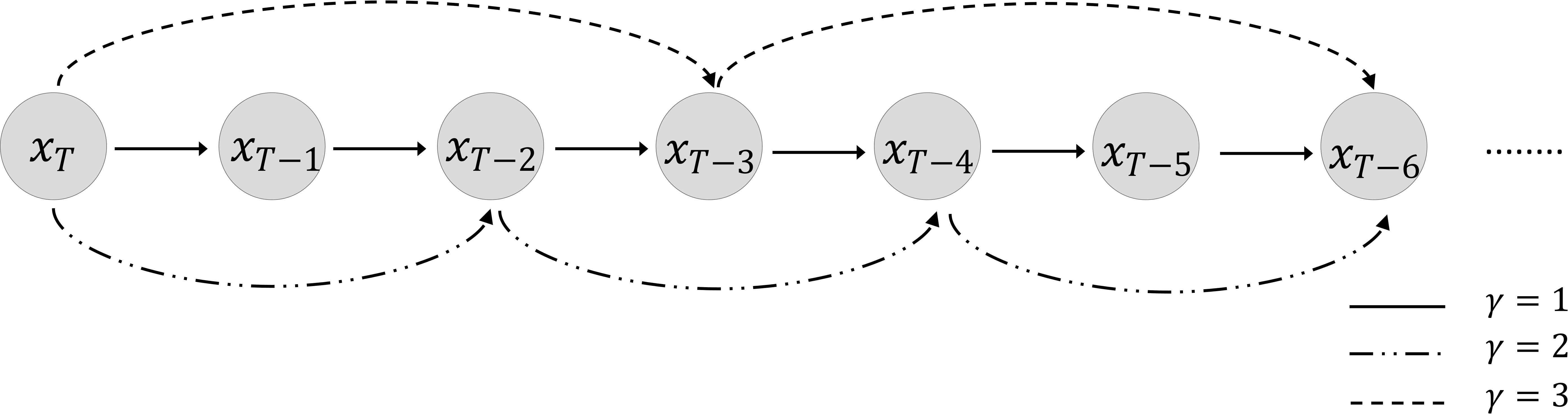}
  \caption{Graphical model for the accelerated sampling with $\gamma=1,2,3$.}
  \label{fig:two diffenrent sampling method}
\end{figure}

\subsection{Accelerated Sampling}
Although the denoising diffusion is a parallel generative model, the inference procedure can take a long time if the number of diffusion steps is large. To mitigate this problem, the denoising diffusion implicit model~(DDIM)~\cite{song2020denoising} introduces a new sampling method called accelerated sampling for diffusion models that boosts up the inference speed without model retraining. The accelerated sampling generates samples over the subsequence of the entire inference trajectory. In particular, it is a quite efficient technique that the sample quality does not deteriorate significantly even with the reduced number of reverse transitions. To improve the synthesis speed while preserving the sample quality, we also leverage the accelerated sampling for Diff-TTS. For implementation, the reverse transition is skipped out by a decimation factor $\gamma$. As shown in Fig.~\ref{fig:two diffenrent sampling method}, the new reverse transitions are composed of periodically selected transitions of the original reverse path. Let $\tau = [\tau_1,\tau_2, ..., \tau_M] (M < T)$ be a new reverse path sampled from the time steps $1,2, ..., T$. The accelerated sampling equation for $i>1$ is described in Eq.~~\eqref{eq7: Diff-TTS accelerated sampling1}:
\begin{equation}
\begin{split}
  x_{\tau_{i-1}} =& \sqrt{\bar{\alpha}_{\tau_{i-1}}}(\frac{x_{\tau_{i}}-\sqrt{1-\bar{\alpha}_{\tau_{i}}}\epsilon_{\theta}(x_{\tau_{i}},\tau_{i},c)}{\sqrt{\bar{\alpha}_{\tau_{i}}}}) \\
  &+ \sqrt{1-\bar{\alpha}_{\tau_{i-1}}-\sigma_{\tau_{i}}^2}\epsilon_{\theta}(x_{\tau_{i}},\tau_{i},c) + \sigma_{\tau_{i}}z_{\tau_{i}},
  \label{eq7: Diff-TTS accelerated sampling1}
 \end{split}
\end{equation}
where $\sigma_{\tau_{i}}=\eta\sqrt{\frac{1-\bar{\alpha}_{\tau_{i-1}}}{1-\bar{\alpha}_{\tau_{i}}}\beta_{\tau_{i}}}$. For $i=1$, the sampling equation follows Eq.~~\eqref{eq7: Diff-TTS accelerated sampling2}:   
\begin{equation}
\begin{split}
x_0=\frac{x_{\tau_1}-\sqrt{1-\bar{\alpha}_{\tau_1}}\epsilon_\theta(x_{\tau_1},\tau_1,c)}{\sqrt{\bar{\alpha}_{\tau_1}}}.
  \label{eq7: Diff-TTS accelerated sampling2}
 \end{split}
\end{equation}
By using the accelerated sampling, Diff-TTS can synthesize the high fidelity mel-spectrogram even sampling over subsequence $\tau$. In other words, the new sampling equation leads to a faster inference speed for the mel-spectrogram generation.

\begin{figure}[t]
  \centering
  \includegraphics[width=\linewidth]{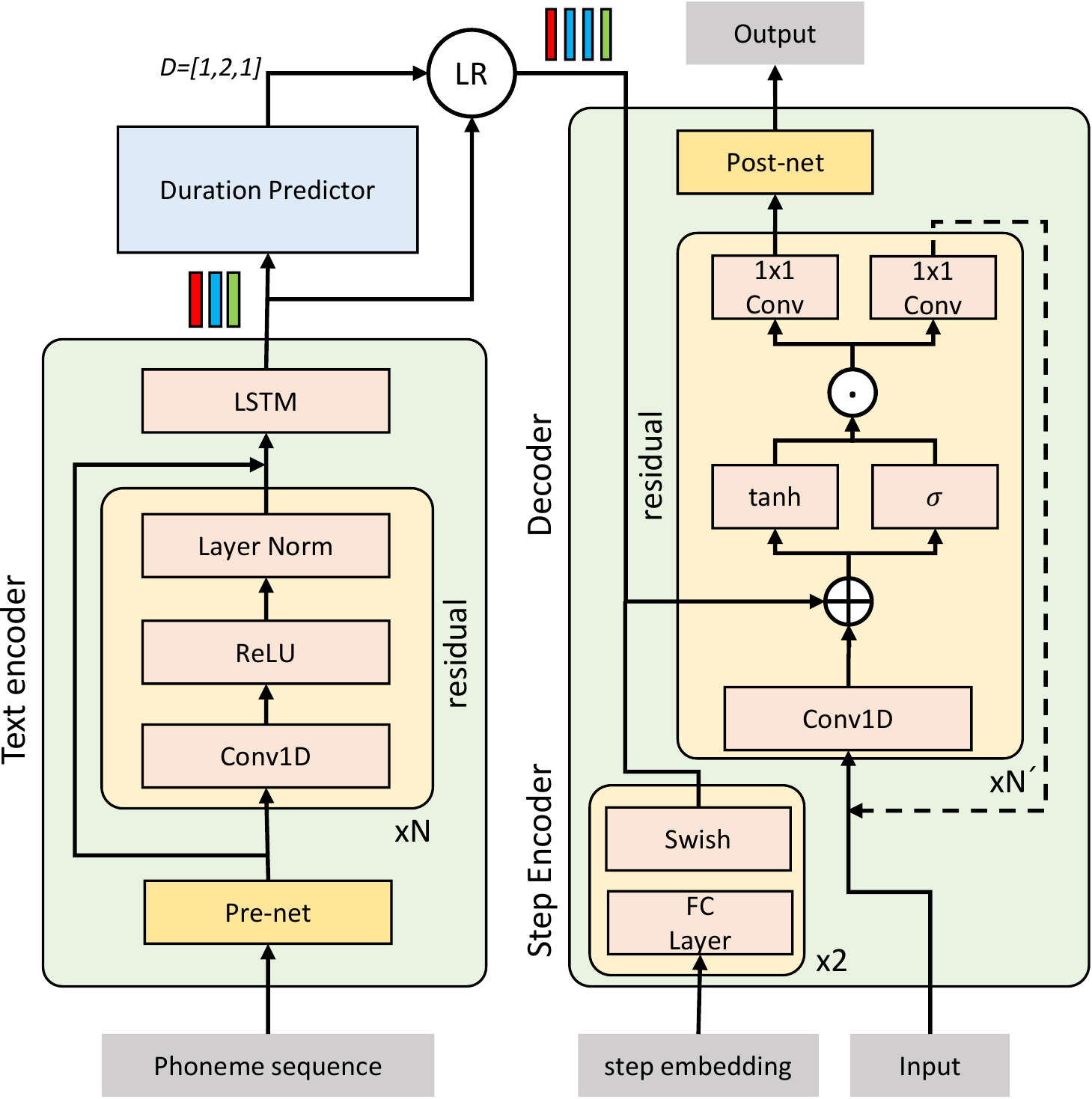}
  \caption{The network architecture of Diff-TTS}
  \label{fig:model}
\end{figure}

\subsection{Model Architecture}
The overall architecture of Diff-TTS is depicted in Fig. 3. Diff-TTS consists of a text encoder, step encoder, duration predictor, and decoder.\\
\textbf{Encoder}: The encoder extracts contextual information from phoneme sequences, then provides it to the duration predictor and the decoder. The encoder part for Diff-TTS is similar to the encoder in~\cite{vainer2020speedyspeech}. The encoder Pre-net starts with embedding and a fully connected (FC) layer with ReLU activation.
Then, the text encoder takes phoneme embedding as input. The encoder module is composed of 10 residual blocks with dilated convolution and an LSTM layer. The dilations of the convolution are [1, 2, 4, 1, 2, 4, 1, 2, 4, 1] from the bottom to the top, and the kernel size is 4 with ReLU activation function followed by layer normalization~\cite{ba2016layer}.\\
\textbf{Duration Predictor and Length Regulator}:
The Diff-TTS employs the length regulator in~\cite{ren2020fastspeech} to match the length of phoneme and mel-spectrogram sequence. The length regulator needs alignment information to expand the phoneme sequence and control the speed of speech. In this paper, Montreal forced alignment~(MFA)~\cite{mcauliffe2017montreal} is used instead of the attention-based alignment extractor, which is commonly used in AR models. The MFA provides more robust alignment than the attention-based alignment and thus improves the alignment accuracy. The duration predictor predicts in the logarithmic domain by using the duration extracted from MFA, which makes the duration prediction stable. The duration predictor is optimized to minimize the L1 loss function.\\
\textbf{Decoder and Step Encoder}: 
In this part, we adapt the similar architecture as in \cite{kong2020diffwave}. The decoder predicts Gaussian noise from the $t$-th step latent variable conditioned on phoneme embedding and diffusion step embedding. The decoder takes the step embedding from a step encoder to be informed about the diffusion time-step so that each diffusion time-step has a different $\epsilon_\theta(\cdot,t,c)$. The diffusion step embedding is a sinusoidal embedding with a 128-dimensional encoding vector for each $t$. The step encoder is constructed using two FC layers and Swish activations~\cite{ramachandran2017searching}. The decoder network consists of a stack of 12 residual blocks with Conv1D, tanh, sigmoid, and 1x1 convolutions with 512 residual channel~\cite{elfwing2018sigmoid}. 
As Figure~\ref{fig:model} shows, the phoneme embedding is expanded by the length regulator. Then, the phoneme embedding and the output of the step encoder are added to the input after the Conv1D layer. The Conv1D layer has a kernel size of 3 without dilation. After going through this residual block, the outputs are summed up before post-net. Finally, the decoder obtains the Gaussian noise corresponding to phoneme sequence and diffusion time-step.

\begin{figure*}%
\centering
\begin{subfigure}{0.9\columnwidth}
\includegraphics[width=\columnwidth]{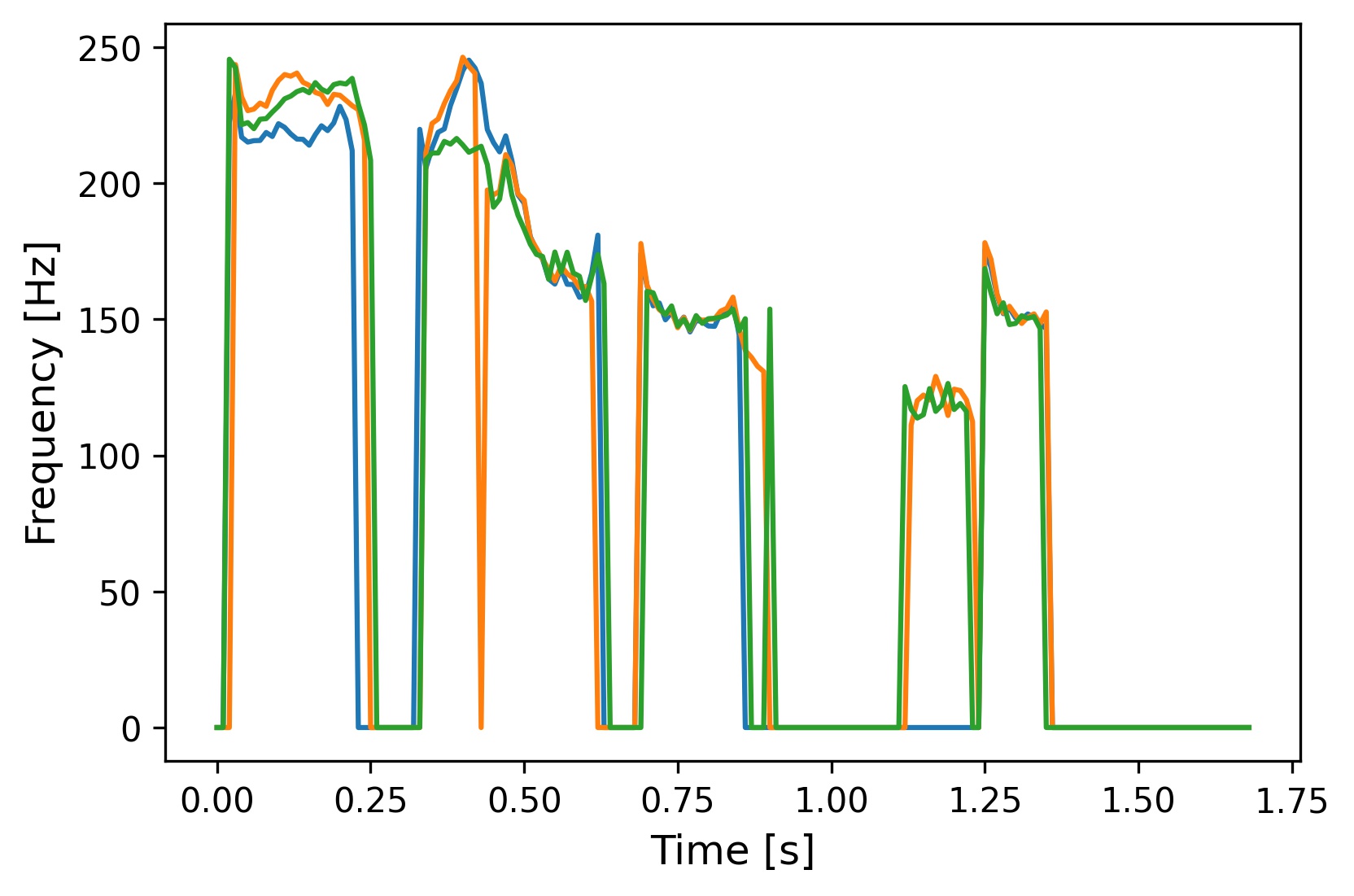}%
\caption{Pitch tracks of the generated speech samples from the same sentence with $\eta=0.2$.}%
\label{subfiga}%
\end{subfigure}\hspace{1cm}
\begin{subfigure}{0.9\columnwidth}
\includegraphics[width=\columnwidth]{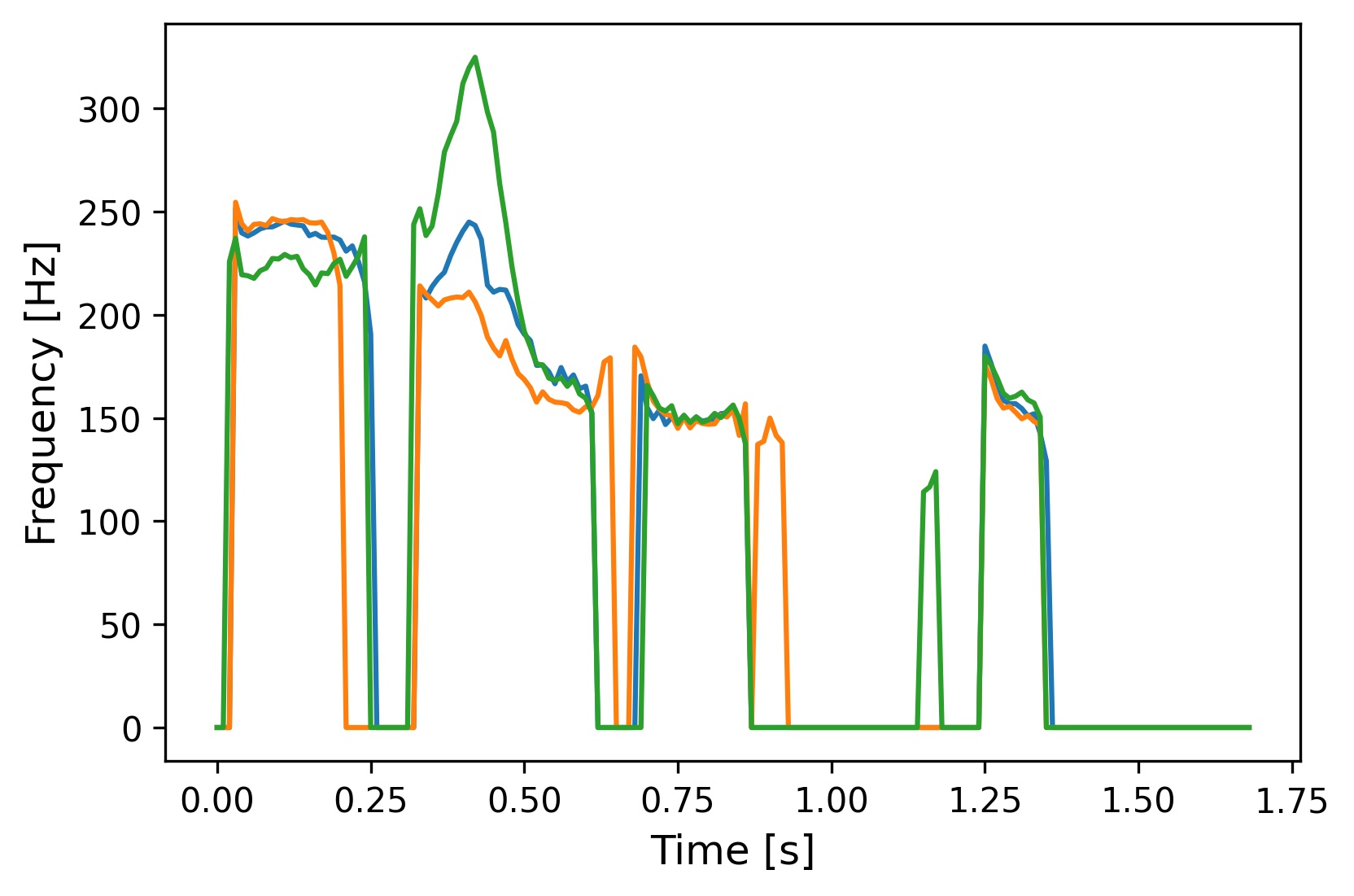}%
\caption{Pitch tracks of the generated speech samples from the same sentence with $\eta=0.6$.}%
\label{subfigb}%
\end{subfigure}
\caption{The fundamental frequency (F0) contours of synthesized speech samples from Diff-TTS.} 
\label{figabc}
\end{figure*}

\section{Experiments}
We evaluated the proposed Diff-TTS with a single female speaker dataset, LJSpeech~\cite{ito2017lj}. It contains 13100 speech samples with a total duration of approximately 24 hours. The dataset was randomly split into the training set~(12,500 samples), validation set~(100 samples), and test set~(500 samples). The Diff-TTS was trained for 700K iterations using Adam optimizer\cite{kingma2014adam} on a single NVIDIA 2080Ti GPU. For quality evaluation, we used pre-trained HiFi-GAN as a neural vocoder that converts mel-spectrogram to the raw waveform.
Our synthesized audio samples\footnote{
https://jmhxxi.github.io/Diff-TTS-demo/index.html.} are publicly available.

\subsection{Audio quality and model size}
\textbf{Audio quality:} 
For the subjective evaluation of audio fidelity, we performed a 5-scale MOS test with 30 audio examples per model and 16 participants. The audio examples were randomly selected from the test dataset. Diff-TTS was trained with 400 time-step and synthesized speech samples with the decimation factor $\gamma\in[1,7,21,57]$. The results are shown in Table~\ref{tab:MOS}. Diff-TTS shows the best performance at the time step of 400 and got an even comparable MOS to the ground truth mel-spectrogram with the same vocoder. On the other hand, the low MOS was obtained with $\gamma=57$, but the degradation is not significant. For any Diff-TTS with accelerated sampling, the performance is as good as Tacotron2 and Glow-TTS. It indicates that accelerated sampling is a practical method without significantly sacrificing speech quality.\\
\textbf{Model size:} 
We compared the number of parameters of Diff-TTS with the Tacotron2 and Glow-TTS. As shown in Table ~\ref{tab:Model size}, Diff-TTS requires the least number of parameters while generating high-fidelity audios. Note that Diff-TTS is about 53\% smaller than Tacotron2 and Glow-TTS. This denotes that the denoising diffusion framework makes Diff-TTS memory-efficient.

\begin{table}[th]
\caption{The Mean Opinion Score~(MOS) of single speaker TTS models with 95\% confidence intervals.}
\label{tab:MOS}
\centering
\begin{tabular}{@{}lc@{}}
\toprule
\textbf{Method}                  & \multicolumn{1}{l}{\textbf{5-scale MOS}} \\ \midrule
GT                               & 4.541 $\pm$ 0.057                                     \\
GT(Mel + HiFiGAN)                & 4.323 $\pm$ 0.060                                     \\
Tacotron2                        & 4.006 $\pm$ 0.072                                  \\
Glow-TTS                         & 4.160 $\pm$ 0.070                                     \\
\midrule
Diff-TTS(T=400, $\gamma=1$)                  & \textbf{4.337 $\pm$ 0.064}                                     \\
Diff-TTS(T=400, $\gamma=7$)                  & 4.223 $\pm$ 0.066                                     \\
Diff-TTS(T=400, $\gamma=21$)                  & 4.135 $\pm$ 0.070                                \\
Diff-TTS(T=400, $\gamma=57$)                  & 4.091 $\pm$ 0.067\\        
\bottomrule
\end{tabular}
\end{table}

\begin{table}[th]
\caption{The comparison of the number of parameters that are used at inference.}
\label{tab:Model size}
\centering
\begin{tabular}{@{}lc@{}}
\toprule
\textbf{Method} & \multicolumn{1}{l}{\textbf{\#of parameters}} \\ \midrule
Tacotron2       & 28.2M                                        \\
Glow-TTS        & 28.6M                                        \\
Diff-TTS        & \textbf{13.4M}                               \\ \bottomrule
\end{tabular}
\end{table}

\subsection{Inference speed}
We measured the average real-time factor~(RTF) to generate mel-spectrograms for 500 sentences of the LJSpeech test set with a single NVIDIA 2080Ti GPU environment. As shown in Table.~\ref{tab:inference speed}, Diff-TTS with $\gamma=57$ achieved an RTF of 0.035 while maintaining higher MOS than Tacotron2. As a comparison, the Tacotron2 model obtained a three times slower inference speed on the same GPU. Although the Diff-TTS is slower than Glow-TTS, Diff-TTS is enough to build a real-time TTS system since it is 28 times faster than real-time. Furthermore, Diff-TTS allows users to control the trade-off between sample quality and inference speed which can be useful when computational resources are limited.

\begin{table}[th]
\caption{The comparison of RTF in mel spectrogram synthesis. RTF denotes the real-time factor, that is the time (in seconds) required for the system to synthesize one second waveform.}
\label{tab:inference speed}
\centering
\begin{tabular}{@{}lcc@{}}
\toprule
\textbf{Method}                & \multicolumn{1}{l}{\textbf{RTF}} \\ \midrule
Tacotron2                & 0.117                                                                             \\ 
Glow-TTS                & \textbf{0.008}                                                                   \\  
\midrule
Diff-TTS(T=400)  \\
\hspace{5mm}Normal diffusion sampling   & 1.744            \\
\hspace{5mm}Accelerated sampling($\gamma$ = 7)                & 0.258                                        \\
\hspace{5mm}Accelerated sampling($\gamma$ = 21)                & 0.090                                                    \\
\hspace{5mm}Accelerated sampling($\gamma$ = 57) & \textbf{0.035}                                                         \\ \bottomrule
\end{tabular}
\end{table}

\subsection{Variability and controllability}
We noticed the prosody of the generated speech depends on the variances of latent representation $x_T$ and Gaussian noise $\sigma_t$ added at the inference. In other words, Diff-TTS can control the prosody by multiplying a temperature term to the variance of latent space and additive noise. To address this further, we illustrated the F0 trajectories of three samples with different temperature terms $\eta \in \{0.2, 0.6\}$. Fig.~\ref{figabc} shows that the larger the temperature term, the more diverse speech is synthesized while maintaining the speech quality. This denotes that Diff-TTS can create many different generative processes without retraining the model, and the variability of prosody can be effectively controlled with the temperature scale.

\section{Conclusions}
In this work, we propose Diff-TTS, which is the first diffusion generative model for a non-AR mel-spectrogram generation. To employ the denoising diffusion for speech synthesis, we present a novel log-likelihood-based optimization method that can synthesize speech appropriate for the text. Then we demonstrate that Diff-TTS achieves controllable and high-fidelity speech synthesis by several experiments. In particular, Diff-TTS has the following advantages: (1) Diff-TTS outperforms Tacotron2 and Glow-TTS in terms of synthesis quality with fewer parameters. (2) By applying accelerated sampling, Diff-TTS allows the user to control the trade-off between sample quality and inference speed. And it has shown that the inference speed of Diff-TTS can be boosted up to 28 times faster than real-time without significantly degrading audio quality. (3) Diff-TTS can control the prosodic variability of speech by multiplying a temperature term. With these advantages, we believe that applying denoising diffusion to speech synthesis can be further developed and refined to produce more realistic waveforms.
For future work, we plan to extend Diff-TTS to support multi-speaker and emotional TTS.

\section{Acknowledgements}
This work was supported by the Institute of Information \& Communications Technology Planning \& Evaluation (IITP) funded by the Korea Government (MSIT) under Grant 2020-0-00059 (Deep learning multi-speaker prosody and emotion cloning technology based on a high quality end-to-end model using small amount of data).

\bibliographystyle{IEEEtran}

\bibliography{mybib}

\begin{thebibliography}{10}
\providecommand{\url}[1]{#1}
\csname url@samestyle\endcsname
\providecommand{\newblock}{\relax}
\providecommand{\bibinfo}[2]{#2}
\providecommand{\BIBentrySTDinterwordspacing}{\spaceskip=0pt\relax}
\providecommand{\BIBentryALTinterwordstretchfactor}{4}
\providecommand{\BIBentryALTinterwordspacing}{\spaceskip=\fontdimen2\font plus
\BIBentryALTinterwordstretchfactor\fontdimen3\font minus
  \fontdimen4\font\relax}
\providecommand{\BIBforeignlanguage}[2]{{%
\expandafter\ifx\csname l@#1\endcsname\relax
\typeout{** WARNING: IEEEtran.bst: No hyphenation pattern has been}%
\typeout{** loaded for the language `#1'. Using the pattern for}%
\typeout{** the default language instead.}%
\else
\language=\csname l@#1\endcsname
\fi
#2}}
\providecommand{\BIBdecl}{\relax}
\BIBdecl

\bibitem{wang2017tacotron}
Y.~Wang, R.~Skerry-Ryan, D.~Stanton, Y.~Wu, R.~J. Weiss, N.~Jaitly, Z.~Yang,
  Y.~Xiao, Z.~Chen, S.~Bengio \emph{et~al.}, ``Tacotron: Towards end-to-end
  speech synthesis,'' \emph{arXiv preprint arXiv:1703.10135}, 2017.

\bibitem{shen2018natural}
J.~Shen, R.~Pang, R.~J. Weiss, M.~Schuster, N.~Jaitly, Z.~Yang, Z.~Chen,
  Y.~Zhang, Y.~Wang, R.~Skerrv-Ryan \emph{et~al.}, ``Natural tts synthesis by
  conditioning wavenet on mel spectrogram predictions,'' in \emph{2018 IEEE
  International Conference on Acoustics, Speech and Signal Processing
  (ICASSP)}.\hskip 1em plus 0.5em minus 0.4em\relax IEEE, 2018, pp. 4779--4783.

\bibitem{ping2017deep}
W.~Ping, K.~Peng, A.~Gibiansky, S.~O. Arik, A.~Kannan, S.~Narang, J.~Raiman,
  and J.~Miller, ``Deep voice 3: Scaling text-to-speech with convolutional
  sequence learning,'' \emph{arXiv preprint arXiv:1710.07654}, 2017.

\bibitem{lee2020bidirectional}
Y.~Lee, J.~Shin, and K.~Jung, ``Bidirectional variational inference for
  non-autoregressive text-to-speech,'' in \emph{International Conference on
  Learning Representations}, 2020.

\bibitem{prenger2019waveglow}
R.~Prenger, R.~Valle, and B.~Catanzaro, ``Waveglow: A flow-based generative
  network for speech synthesis,'' in \emph{ICASSP 2019-2019 IEEE International
  Conference on Acoustics, Speech and Signal Processing (ICASSP)}.\hskip 1em
  plus 0.5em minus 0.4em\relax IEEE, 2019, pp. 3617--3621.

\bibitem{kong2020hifi}
J.~Kong, J.~Kim, and J.~Bae, ``Hifi-gan: Generative adversarial networks for
  efficient and high fidelity speech synthesis,'' \emph{arXiv preprint
  arXiv:2010.05646}, 2020.

\bibitem{oord2016wavenet}
A.~v.~d. Oord, S.~Dieleman, H.~Zen, K.~Simonyan, O.~Vinyals, A.~Graves,
  N.~Kalchbrenner, A.~Senior, and K.~Kavukcuoglu, ``Wavenet: A generative model
  for raw audio,'' \emph{arXiv preprint arXiv:1609.03499}, 2016.

\bibitem{ping2018clarinet}
W.~Ping, K.~Peng, and J.~Chen, ``Clarinet: Parallel wave generation in
  end-to-end text-to-speech,'' \emph{arXiv preprint arXiv:1807.07281}, 2018.

\bibitem{li2019neural}
N.~Li, S.~Liu, Y.~Liu, S.~Zhao, and M.~Liu, ``Neural speech synthesis with
  transformer network,'' in \emph{Proceedings of the AAAI Conference on
  Artificial Intelligence}, vol.~33, no.~01, 2019, pp. 6706--6713.

\bibitem{valle2020flowtron}
R.~Valle, K.~Shih, R.~Prenger, and B.~Catanzaro, ``Flowtron: an autoregressive
  flow-based generative network for text-to-speech synthesis,'' \emph{arXiv
  preprint arXiv:2005.05957}, 2020.

\bibitem{peng2020non}
K.~Peng, W.~Ping, Z.~Song, and K.~Zhao, ``Non-autoregressive neural
  text-to-speech,'' in \emph{International Conference on Machine
  Learning}.\hskip 1em plus 0.5em minus 0.4em\relax PMLR, 2020, pp. 7586--7598.

\bibitem{lancucki2020fastpitch}
A.~{\L}a{\'n}cucki, ``Fastpitch: Parallel text-to-speech with pitch
  prediction,'' \emph{arXiv preprint arXiv:2006.06873}, 2020.

\bibitem{miao2020efficienttts}
C.~Miao, S.~Liang, Z.~Liu, M.~Chen, J.~Ma, S.~Wang, and J.~Xiao,
  ``Efficienttts: An efficient and high-quality text-to-speech architecture,''
  \emph{arXiv preprint arXiv:2012.03500}, 2020.

\bibitem{donahue2020end}
J.~Donahue, S.~Dieleman, M.~Bi{\'n}kowski, E.~Elsen, and K.~Simonyan,
  ``End-to-end adversarial text-to-speech,'' \emph{arXiv preprint
  arXiv:2006.03575}, 2020.

\bibitem{ren2019fastspeech}
Y.~Ren, Y.~Ruan, X.~Tan, T.~Qin, S.~Zhao, Z.~Zhao, and T.-Y. Liu, ``Fastspeech:
  Fast, robust and controllable text to speech,'' \emph{arXiv preprint
  arXiv:1905.09263}, 2019.

\bibitem{ren2020fastspeech}
Y.~Ren, C.~Hu, T.~Qin, S.~Zhao, Z.~Zhao, and T.-Y. Liu, ``Fastspeech 2: Fast
  and high-quality end-to-end text-to-speech,'' \emph{arXiv preprint
  arXiv:2006.04558}, 2020.

\bibitem{vainer2020speedyspeech}
J.~Vainer and O.~Du{\v{s}}ek, ``Speedyspeech: Efficient neural speech
  synthesis,'' \emph{arXiv preprint arXiv:2008.03802}, 2020.

\bibitem{miao2020flow}
C.~Miao, S.~Liang, M.~Chen, J.~Ma, S.~Wang, and J.~Xiao, ``Flow-tts: A
  non-autoregressive network for text to speech based on flow,'' in
  \emph{ICASSP 2020-2020 IEEE International Conference on Acoustics, Speech and
  Signal Processing (ICASSP)}.\hskip 1em plus 0.5em minus 0.4em\relax IEEE,
  2020, pp. 7209--7213.

\bibitem{kim2020glow}
J.~Kim, S.~Kim, J.~Kong, and S.~Yoon, ``Glow-tts: A generative flow for
  text-to-speech via monotonic alignment search,'' \emph{arXiv preprint
  arXiv:2005.11129}, 2020.

\bibitem{ho2020denoising}
J.~Ho, A.~Jain, and P.~Abbeel, ``Denoising diffusion probabilistic models,''
  \emph{arXiv preprint arXiv:2006.11239}, 2020.

\bibitem{xiao2020vaebm}
Z.~Xiao, K.~Kreis, J.~Kautz, and A.~Vahdat, ``Vaebm: A symbiosis between
  variational autoencoders and energy-based models,'' \emph{arXiv preprint
  arXiv:2010.00654}, 2020.

\bibitem{nichol2021improved}
A.~Nichol and P.~Dhariwal, ``Improved denoising diffusion probabilistic
  models,'' \emph{arXiv preprint arXiv:2102.09672}, 2021.

\bibitem{chen2020wavegrad}
N.~Chen, Y.~Zhang, H.~Zen, R.~J. Weiss, M.~Norouzi, and W.~Chan, ``Wavegrad:
  Estimating gradients for waveform generation,'' \emph{arXiv preprint
  arXiv:2009.00713}, 2020.

\bibitem{kong2020diffwave}
Z.~Kong, W.~Ping, J.~Huang, K.~Zhao, and B.~Catanzaro, ``Diffwave: A versatile
  diffusion model for audio synthesis,'' \emph{arXiv preprint
  arXiv:2009.09761}, 2020.

\bibitem{song2020denoising}
J.~Song, C.~Meng, and S.~Ermon, ``Denoising diffusion implicit models,''
  \emph{arXiv preprint arXiv:2010.02502}, 2020.

\bibitem{ba2016layer}
J.~L. Ba, J.~R. Kiros, and G.~E. Hinton, ``Layer normalization,'' \emph{arXiv
  preprint arXiv:1607.06450}, 2016.

\bibitem{mcauliffe2017montreal}
M.~McAuliffe, M.~Socolof, S.~Mihuc, M.~Wagner, and M.~Sonderegger, ``Montreal
  forced aligner: Trainable text-speech alignment using kaldi.'' in
  \emph{Interspeech}, vol. 2017, 2017, pp. 498--502.

\bibitem{ramachandran2017searching}
P.~Ramachandran, B.~Zoph, and Q.~V. Le, ``Searching for activation functions,''
  \emph{arXiv preprint arXiv:1710.05941}, 2017.

\bibitem{elfwing2018sigmoid}
S.~Elfwing, E.~Uchibe, and K.~Doya, ``Sigmoid-weighted linear units for neural
  network function approximation in reinforcement learning,'' \emph{Neural
  Networks}, vol. 107, pp. 3--11, 2018.

\bibitem{ito2017lj}
K.~Ito and L.~Johnson, ``The lj speech dataset,'' \emph{Online:
  https://keithito. com/LJ-Speech-Dataset}, 2017.

\bibitem{kingma2014adam}
D.~P. Kingma and J.~Ba, ``Adam: A method for stochastic optimization,''
  \emph{arXiv preprint arXiv:1412.6980}, 2014.

\end{thebibliography}

\end{document}